\documentclass[superscriptaddress,reprint,amsmath,amssymb,aps,
floatfix,
]{revtex4-1}

\usepackage{graphicx}
\usepackage{dcolumn}
\usepackage{bm,MnSymbol}

\newcommand{\bra}[1]{\langle #1 \vert}
\newcommand{\ket}[1]{\vert #1 \rangle}

\begin{document}

\title{Experimental test of photonic entanglement in accelerated reference frames}
\thanks{Correspondence and requests for materials should be addressed to
 Matthias Fink and Rupert Ursin.}

\author{Matthias Fink}
\email{Matthias.fink@oeaw.ac.at}
\affiliation{Institute for Quantum Optics and Quantum Information - Vienna (IQOQI), Austrian Academy of Sciences, Vienna, Austria}
\author{Ana Rodriguez-Aramendia}
\affiliation{Institute for Quantum Optics and Quantum Information - Vienna (IQOQI), Austrian Academy of Sciences, Vienna, Austria}
\author{Johannes Handsteiner}
\affiliation{Institute for Quantum Optics and Quantum Information - Vienna (IQOQI), Austrian Academy of Sciences, Vienna, Austria}
\author{Abdul Ziarkash}
\affiliation{Institute for Quantum Optics and Quantum Information - Vienna (IQOQI), Austrian Academy of Sciences, Vienna, Austria}
\author{Fabian Steinlechner}
\affiliation{Institute for Quantum Optics and Quantum Information - Vienna (IQOQI), Austrian Academy of Sciences, Vienna, Austria}
\author{Thomas Scheidl}
\affiliation{Institute for Quantum Optics and Quantum Information - Vienna (IQOQI), Austrian Academy of Sciences, Vienna, Austria}
\author{Ivette Fuentes}
\affiliation{Faculty of Physics, University of Vienna, Boltzmanngasse 5, A-1090 Vienna, Austria}
\author{Jacques Pienaar}
\affiliation{Institute for Quantum Optics and Quantum Information - Vienna (IQOQI), Austrian Academy of Sciences, Vienna, Austria}
\affiliation{Faculty of Physics, University of Vienna, Boltzmanngasse 5, A-1090 Vienna, Austria}
\author{Tim C. Ralph}
\affiliation{Centre for Quantum Computation \& Communication Technology, Department of Physics, University of Queensland, Australia}
\author{Rupert Ursin}
\email{Rupert.Ursin@oeaw.ac.at}
\affiliation{Institute for Quantum Optics and Quantum Information - Vienna (IQOQI), Austrian Academy of Sciences, Vienna, Austria}
\affiliation{Vienna Center for Quantum Science and Technology (VCQ), Vienna, Austria}

\date{\today}

\begin{abstract}
The quantization of the electromagnetic field has successfully paved the way for the development of the Standard Model of Particle Physics and has established the basis for quantum technologies. Gravity, however, continues to hold out against physicists' efforts of including it into the framework of quantum theory. Experimental techniques in quantum optics have only recently reached the precision and maturity required for the investigation of quantum systems under the influence of gravitational fields. Here, we report on experiments in which a genuine quantum state of an entangled photon pair was exposed to a series of different accelerations. We measure an entanglement witness for $g$ values ranging from $30 mg$ to up to $30 g$ - under free-fall as well on a spinning centrifuge - and have thus derived an upper bound on the effects of uniform acceleration on photonic entanglement. Our work represents the first quantum optics experiment in which entanglement is systematically tested in geodesic motion as well as in accelerated reference frames with acceleration $a>>g = 9{,}81 \frac{m}{s^2}$.
\end{abstract}

\maketitle

\section{\label{sec:Quantum}Quantum Theory vs. Relativity}

Einstein's Relativity Theory (RT) and quantum theory (QT) are both important pillars of modern physics and have been thoroughly tested by many high precision experiments. RT has proven its ability to describe the Universe on the largest scales. Conversely, QT was invented to describe the Universe at atomic and subatomic scales. All currently known physical effects can be described by either the one or the other theory to very high precision. Without the scientific insights of these two theories, a modern society would not have been imaginable, taking the global positioning system (GPS) or the importance of semiconductors as two prominent examples. 

In a broader view, the unification of Einstein’s theory of relativity and quantum mechanics is a long-standing problem in contemporary physics. As long as the existing descriptions of nature remain confined to their own specialized regimes, they cannot contribute a unified theory that captures physics at the boundary between these regimes. Current models are therefore considered incomplete, and can at best be combined in a patchwork way, and only under special conditions. In natural science, where theory becomes uncertain, experimental guidance is needed. It is not unusual for the first steps into a new physical regime to turn up unexpected experimental results, as occurred in the historical black-body radiation \cite{Rubens:1901} and Michelson-Morley \cite{Michelson:1887} experiments. 

Dimensional arguments suggest that quantum gravity must dominate close to the Planck scale, which is still far beyond current experimental capabilities. However, there remains the distinct possibility that quantum gravity phenomena already become significant at scales that are just becoming accessible. Quantum entangled systems subjected to high and low accelerations is one regime where new physical phenomena might be expected to arise. Also, experimental investigations involving hyper- or microgravity can cause unexpected changes to physical phenomena \cite{Dreyer:2010}. Exposing physical systems to such extreme conditions can aid in the understanding of that system, and lead to a deeper understanding of the physical processes themselves.

It has been conjectured e.g. by Penrose \cite{Penrose:1996ty} and Diosi \cite{diosi1987universal} that gravity may cause a quantum state to collapse. This is just one of many ways that researchers have tried to include gravity into the quantum theory paradigm \cite{Milburn:1991tn, Adler:2000, Ralph:2014, Pikovski:2015, Blencowe:2013}. Gravity and acceleration can have observable effects on quantum entanglement \cite{Bruschi2014a, Bruschi2012, Friis2012, Alsing2012}. However, to date there has been no systematic experimental investigation of what happens to quantum entanglement under various gravitational fields and accelerations. With the onset of new technology, techniques now exist to transgress experimental barriers and begin to test quantum physics in regimes where gravity is important. However, experimental testing of quantum phenomena in these regimes is still sparse \cite{Amelino-Camelia2014}. To date, first experiments have been performed using only single quanta, e.g. in the pioneering work of Colella et al in 1975 \cite{Colella:1975, Abele:2012} and correlated photon pairs in microgravity \cite{Tang:2016}. Just recently a remarkable one-shot experiment was carried out where an entangled photon source was exposed to explosive g-forces, but not during operation \cite{Tang:2016}.

Following these lines we report on a series of experiments in which a provably quantum state \cite{Bell:1964} of photon pairs entangled in their polarization degree of freedom is exposed to uniform accelerations. The acceleration is imposed by its motion, in free-fall as well as on a centrifuge.

Apart from the inherent novelty of testing entanglement under different accelerated conditions, the results of the experiment can also be extrapolated to analogous hypothetical experiments under different gravitational fields. If, for example, the equivalence principle holds, then a local test of entanglement should not reveal the difference between uniform acceleration and a gravitational field \cite{Zych:2015}. The experiment therefore implies bounds on the amount of decoherence as well as potential unexpected unitary transformations caused by gravity, such as in the speculative models mentioned above. Our experiment takes a broad-brush approach by aiming to establish a bound on any arbitrary effect of this kind, without focusing on any specific model.

Firstly, a crate containing an entangled photon source and two single-photon polarization detection units was dropped from $12m$ in a drop tower to realize a microgravity environment, where we reached $30$ $mg$. Secondly, that crate was mounted on an arm of a rotational centrifuge and accelerated to as high as $30$ $g$ (comparable to the surface gravity of the sun) at a maximum angular speed of  $9{.}9$ $rad$ $sec^{-1}$. We evaluated the impact of the acceleration on that genuine quantum system, by constantly monitoring an entanglement witness. Our results show (assuming the equivalence principle holds) that quantum entanglement is unaffected by ambient gravitational conditions to within the resolution of our test-system. The gravity experienced by the entangled state ranges from those on the orbiting ISS, Mars, Jupiter or close to the sun\footnote{Of course, these environments would pose other challenges to entanglement, such as exposure to radiation and strong magnetic fields, which we do not address here.}. This represents the first experimental effort exposing a genuine quantum system to micro- and hyper-gravitation and extends the experimental regime in which quantum effects can be said to exist in harmony with gravity. Future experiments could be designed to show that entanglement changes when the system undergoes non-uniform acceleration. Such experiments would require much higher accelerations or massive fields.

\section{\label{sec:Methods}Entangled photon source} 

\begin{figure*}\label{source}
\includegraphics[width=0.39\textwidth]{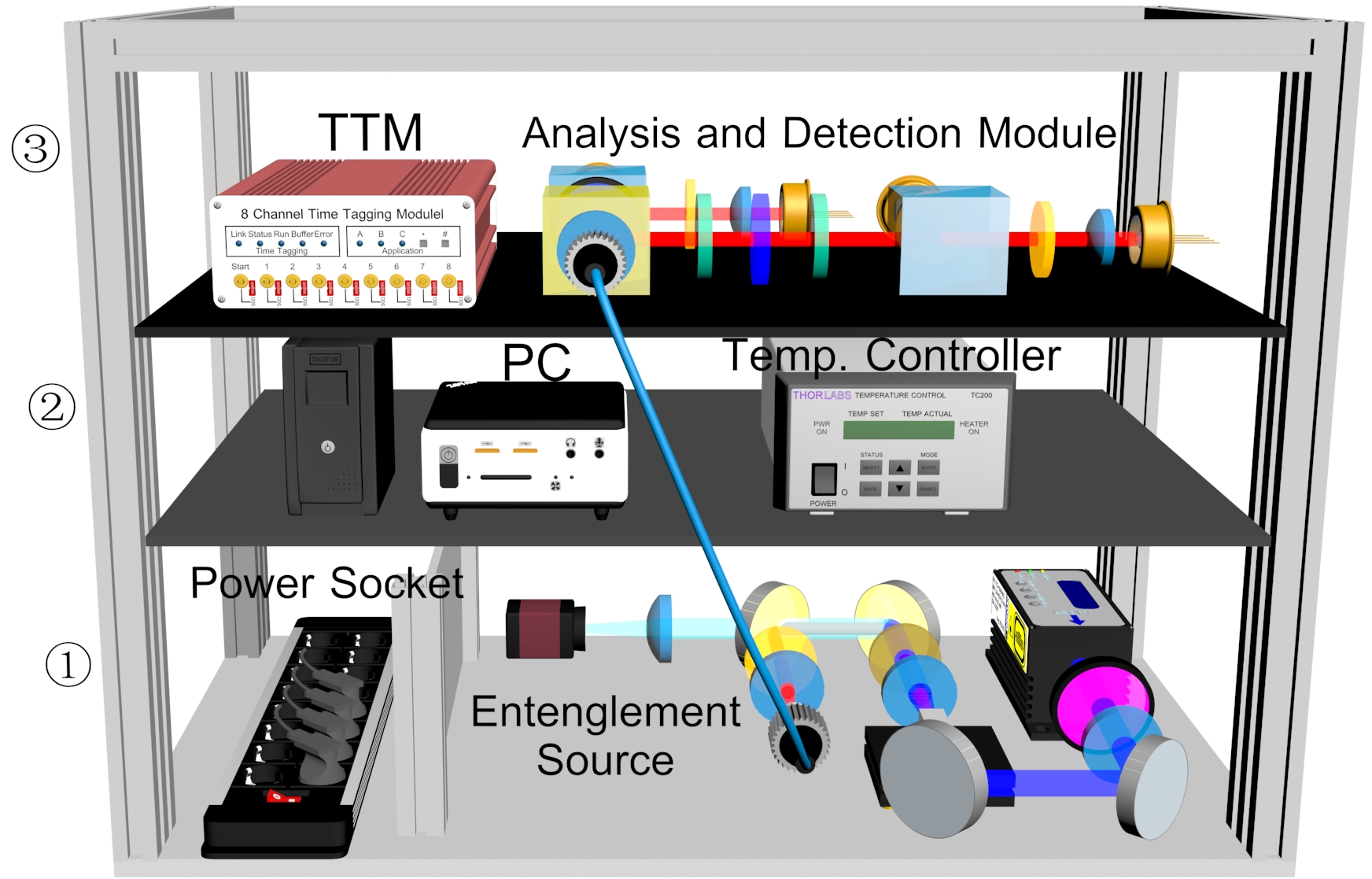} 
\includegraphics[width=0.6\textwidth]{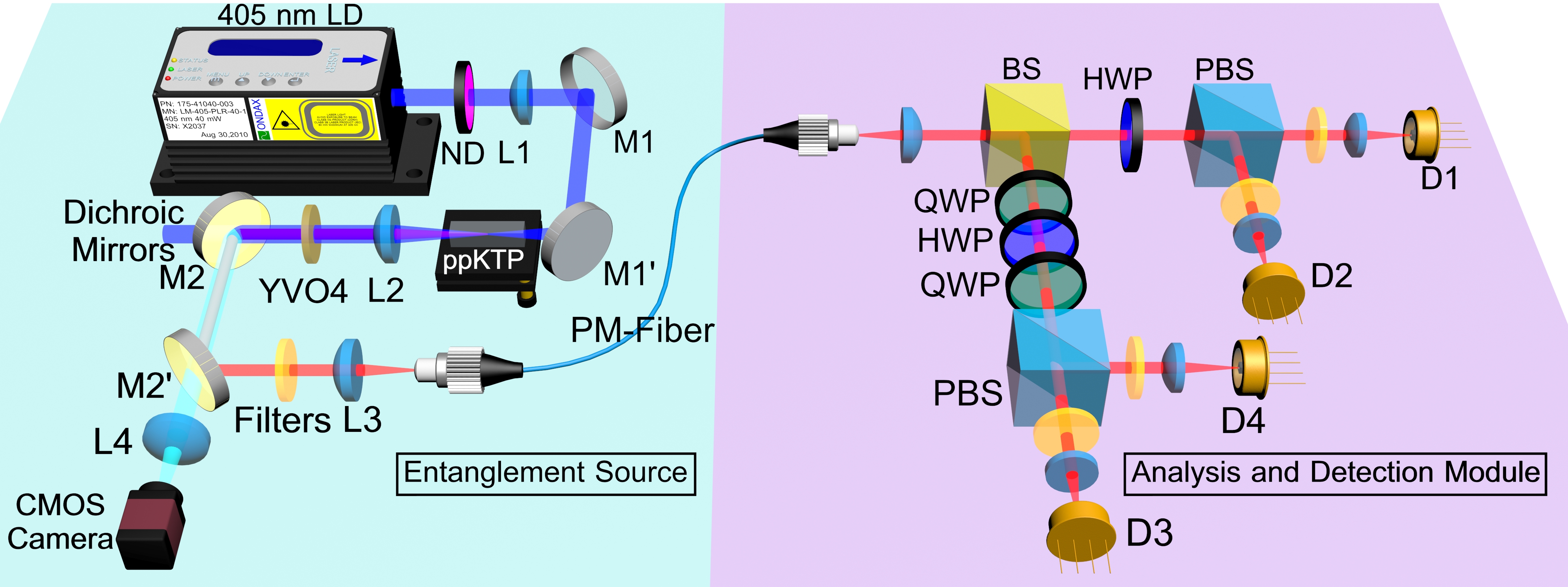} 
\caption{Sketch of the source crate used in the experiments. Horizontal projection right: A laser diode (LD) pumps a ppKTP crystal heated by an oven, generating photon pairs collected in a polarization maintaining single-mode fiber. An additional Nd:YVO4 crystal is used to fine-tune the walk-off compensation. A beam-splitter (BS) randomly splits the photon pairs, thus yielding a polarization-entangled state (in postselection). Using various half- and quarter-wave plates (HWP, QWP) and polarizing beam-splitters (PBS), the polarization correlations can be analyzed in different measurement bases and measured in detectors (D1,D2,D3 and D4). Vertical projection left: Source (level 3), electronics (level 2) as well as polarization analysis and detection module (level 1) are placed at different levels inside the crate and thus exhibit different maximal g-values.}
\end{figure*}

\begin{figure*}\label{towercentri}
\includegraphics[width=0.29\textwidth]{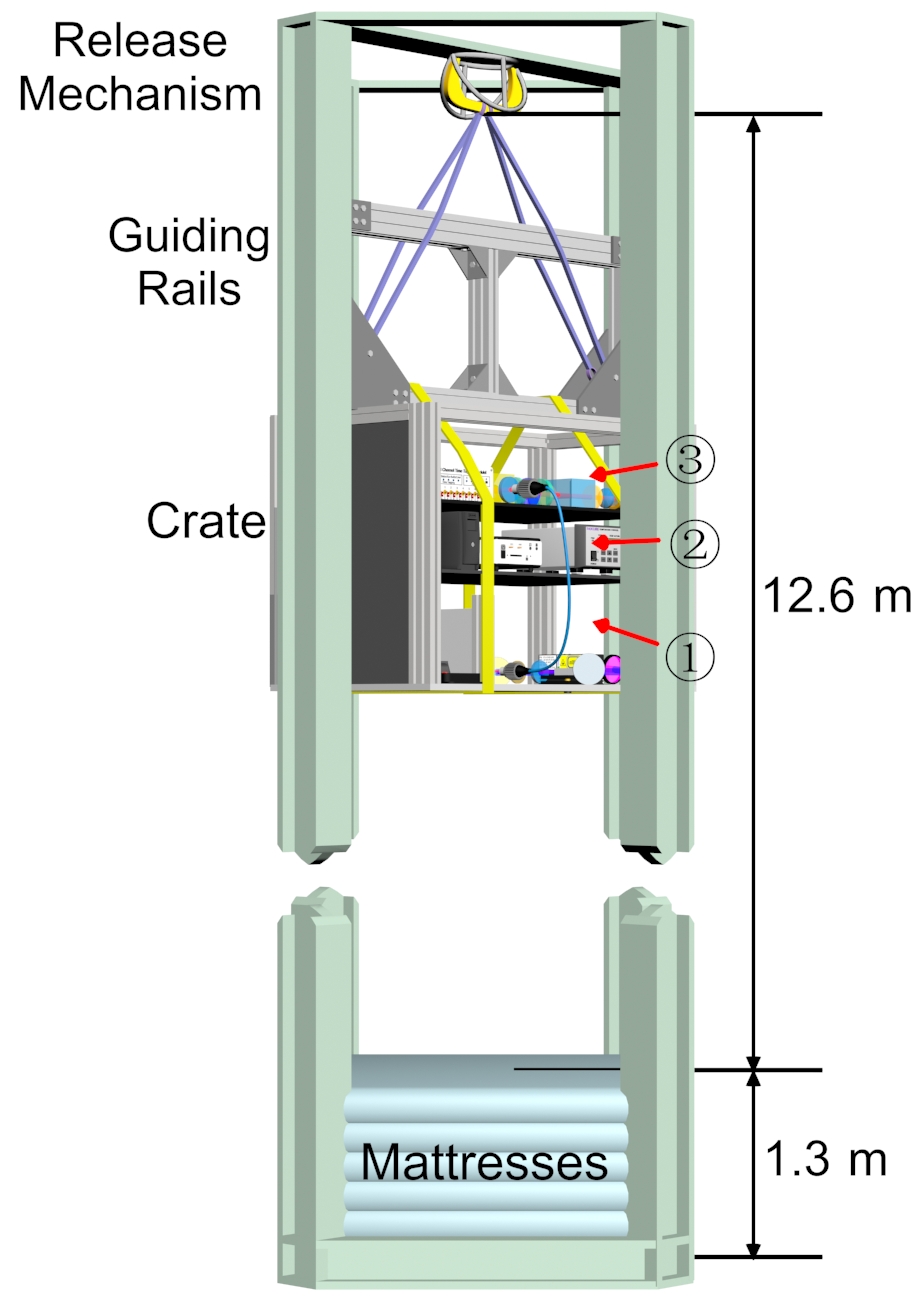}
\includegraphics[width=0.69\textwidth]{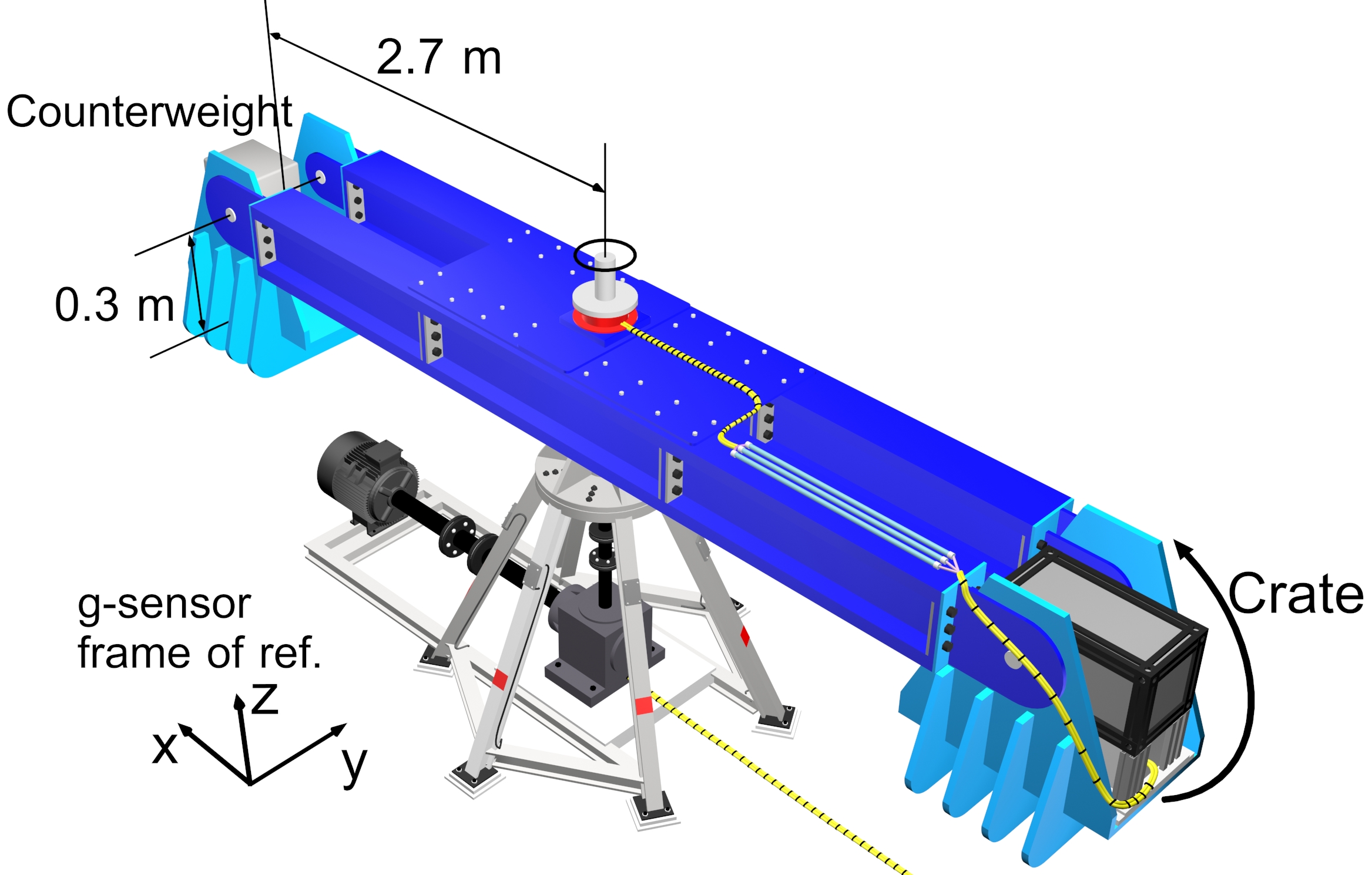}
 \llap{\raisebox{4cm}[0\textwidth][0\textwidth]{
    \includegraphics[height=3cm]{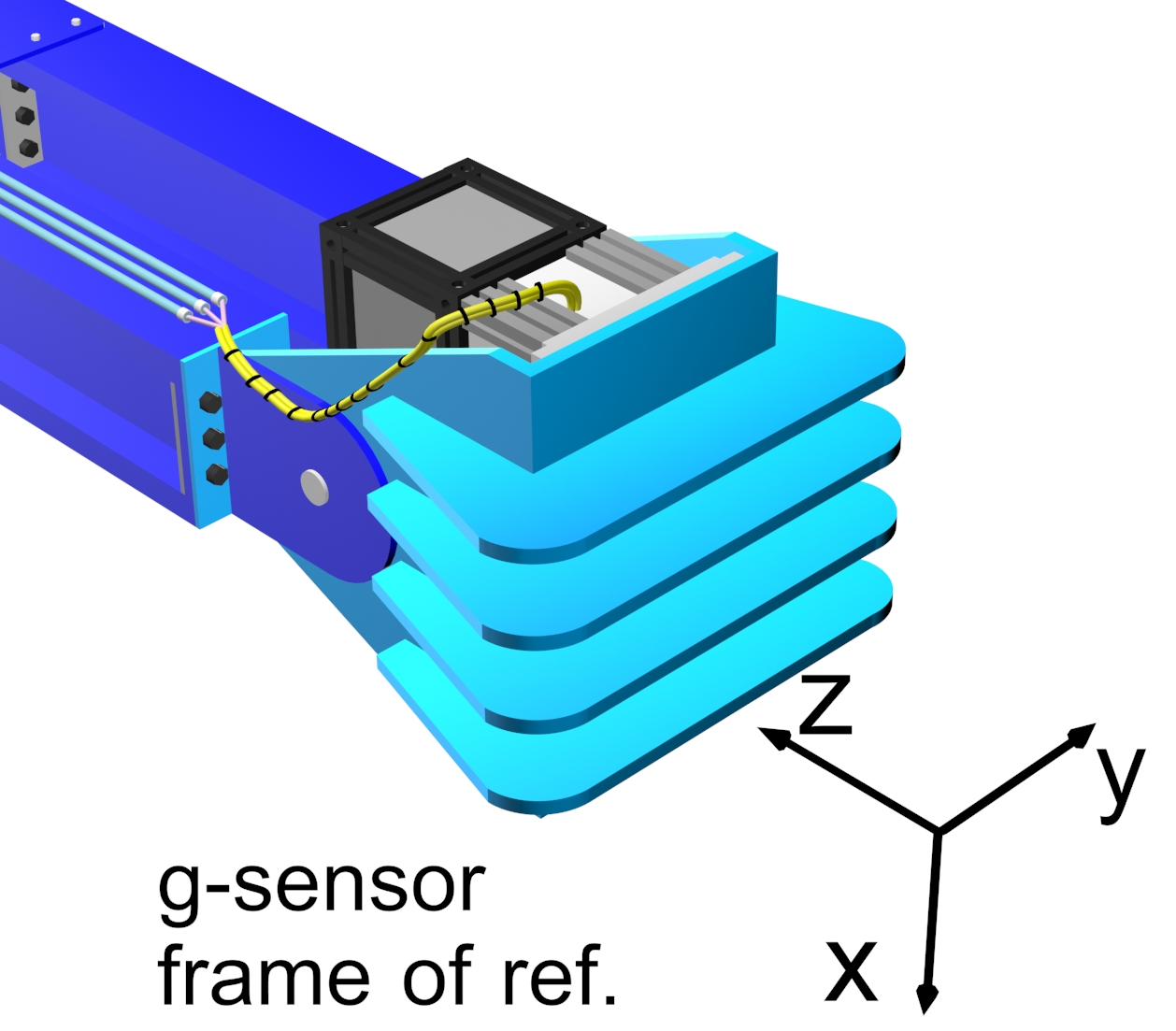}
    }
    }
\caption{Left: A sketch of the drop tower experiment. The crate was dropped from $12$ $m$, for a low-g (almost) free-fall flight in air, lasting $1{.}4$ $sec$. A $1{.}3$ $m$ high stack of mattresses was used to minimize impact of the crate deceleration (see supplementary videos Sub\_1.mp4). Right: The centrifuge used for the experiments. The source-crate was embedded in one of the centrifuge gondolas at a distance of about $3$ $m$ from the axis, at full speed. The orientations of the accelerometer frame of reference when the centrifuge is at rest (a) and rotating (b) are shown (see supplementary videos Sub\_2.mp4).}
\end{figure*}

In order to guarantee the stability requirements for our intended drop-tower and centrifuge experiments, we designed a simple and very rigid entangled photon source capable of withstanding the high g-forces during the deceleration and acceleration phases (see Fig \ref{source}) of the experimental runs. The source (see center and right part of Fig. \ref{source}) is based on spontaneous parametric down conversion (SPDC) in a periodically poled Potassium Titanyl Phosphate (ppKTP) crystal in a degenerate collinear type-II quasi-phase matching configuration and post-selection on a beam splitter (BS) \cite{Kuklewicz:2004}. A continuous-wave pump laser at $405$ $nm$ is focused loosely into the crystal and creates pairs of photons at $810$ $nm$, with horizontally  and vertically polarized  signal and idler photons, respectively. The crystal is temperature stabilized to $39{.}6^{\circ}C \pm 0{.}1^{\circ}C$, to ensure degenerate phase matching \cite{Kiess:1993wn}. After being separated from the pump light by two dichoric mirrors and spectral filtered using an interference filter, the two photons are coupled into a sole polarization maintaining single-mode optical fiber (PMSMF). In order to monitor possible misalignment of the beam paths during acceleration, a fraction of the pump beam was imaged onto a CMOS camera that was placed behind one of the dichroic mirrors. The length of the PMSMF and an additional neodymium-doped yttrium orthovanadate (Nd:YVO4) crystal were chosen, such that the average longitudinal walk-off between the signal and idler photons due to the birefringence of the ppKTP crystal, was compensated. This ensures, that no information about the polarization of a photon can be obtained from the timing of the SPDC photons. The PMSMF also had the benefit of ensuring polarization stability even under high accelerations during an experimental run. The PMSMF guides the photons to a beam splitter, where they are split with $50\%$ probability. As a consequence, one obtains the maximally polarization entangled $\ket{\Psi^+}=1/\sqrt{2}(\ket{H_1V_2}+\ket{V_1H_2})$ state for the photon pairs that are split up at the BS in mode $1$ and $2$. Here, $H$ ($V$) denote the horizontal (vertical) polarization, where we define the linear polarization state relative to the base plate of the source, as it's orientation changes during the experiment (Fig. \ref{towercentri} right). Note that the orientation of this reference frame changes during the experiment relative to the laboratory frame.

A combination of a quarter-, half- and quarter-wave plate (QHQ) is used to compensate polarization dependent phase shifts caused by the reflection at the BS and allows setting different Bell states. For the presented experiments we set the QHQ such that we end up with an rotational invariant two photon state $\ket{\Psi^-}=1/\sqrt{2}(\ket{H_1V_2}-\ket{V_1H_2})$. An additional motorized half-wave plate followed by a polarizing beam splitter (PBS) is inserted in each output mode of the 50/50 BS to analyze the polarization of the photons in any desired linear polarization measurement basis. Finally, the photons were detected using four passively quenched semiconductor avalanche photo diodes placed in every output mode of the two PBS. Colored glass and additional interference filters were placed in front of the detectors in order to minimize background counts from the remaining pump photons and stray background light. The electronic detector signals are recorded by a time-tagging module, allowing to post-select simultaneous detection events in the two output modes of the 50/50 BS. The optical setup together with the required electronic equipment was confined in a rigid 3-level crate (Fig. \ref{source} left), constituting a fully functional and ultra-stable entangled-photon box. This facilitated flexible plug-and-play installation at the drop-tower and centrifuge support structures.

The Bell-state fidelity of the experimental state generated at various g values was assessed by measuring a fidelity witness $F_{\Psi^-}(\hat{\rho}_{exp}) = \bra{\Psi^-} \hat{\rho}_{exp} \ket{\Psi^-}$ with $F_{\Psi^-}(\hat{\rho}_{exp}) \geq F_{\Psi^-}^M (\hat{\rho}_{exp}) =  \frac{1}{2}(V_{HV}+V_{DA})$ \cite{Blinov:2004}. Here, $V_{HV}$ and $V_{DA}$ denote the visibilities measured in the horizontal/vertical (diagonal/anti-diagonal) basis $V=\frac{N_{12}+N_{21}-N_{11}-N_{22}}{N_{12}+N_{21}+N_{11}+N_{22}}$, where $N$ is the number of coincidence detection events between detectors 1 and 2 for each measurement setting (HV,DA). For a pump power of approximately $5$ $mW$ The setup provides $280$ $kcps$ detected single counts in total and $7$ $kcps$ coincident counts, yielding a visibility of $V_{HV}=97\%$ in HV-basis and $V_{DA}=96\%$ in the DA basis. Note that the visibility in the DA basis is lower, as it is affected by changes in the polarization basis measured as well as the partial distinguishably of the two photons (non-degenerate phase matching).

\section{\label{sec:drop}The drop tower experiment} 
A $12$ $m$ high drop was used for the micro-g experiment which provides us with $1{.}4$ $sec$ of integration time (see Fig. \ref{towercentri}). Power supply was provided by a battery built into the crate capable to keep the source and detectors alive for about $1{.}5$ $h$, a wireless network antenna provided data connection from the control room to the falling crate. Guide rails are used to keep the crate on track during flight, so as not to be displaced by Coriolis-force due to the earth's rotation during the free-fall phase. Wind speed and the guide rails resulted in drag on the crate, which reaches $55$ $km/h$ of maximum speed, enabling us to measure down to $30 \pm 3$ $mg$ after the solenoid based release mechanism was opened. For a smooth deceleration, a high stack of foam mattresses was used, which still caused a saturation of our g-sensor at $16$ $g$. 

\begin{figure*}
\includegraphics[width=0.8\textwidth]{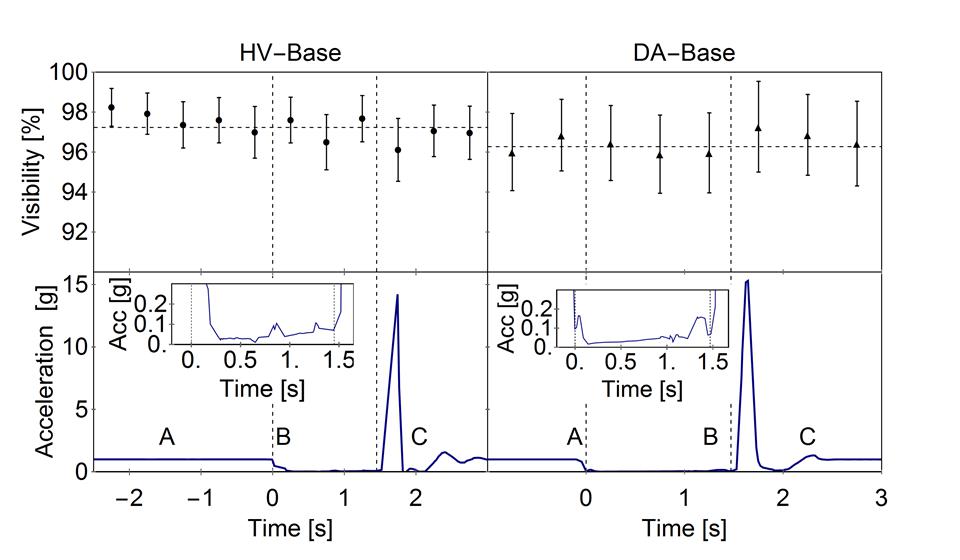}
\caption{\label{g:drop}Droptower data: Elapsed time at rest (A), free-fall flight (B) and after impact (C) versus HV (left) and DA-Visibility (right) measured in $500$ $ms$ slots compared to the its average (denoted as horizontal dashed line). Each of data points shown consists of $\approx 3{,}500$ coincident counts and $V_{DA}=96\%$ on average. The actual g-value shows the drag from a few strikes from the guiding rail and the wind at higher velocities. The impact phase (at $1{.}45$ sec), shows $16$ $g$ deceleration (g-sensor saturated) and rebound (at $2$ sec).}
\end{figure*}

We observed no lasting degradation in count rates or source visibility after impact on the mattresses, and could repeat the experiment many times without the need for re-alignment between measurement runs. In each successive experimental run we measured either in the HV- or in the DA-visibility shown in Fig \ref{g:drop}, which gives $\approx 400{,}000$ single counts in total stored as time-tags  on the local computer for later evaluation. During the free-fall phase the visibility never dropped below $V_{DA}=96\% \pm 2\%$.

\section{\label{sec:level1}The centrifuge experiment} 
For the hyper-gravity regime, we used a centrifuge that puts the crate in rotation around a fixed vertical axis, applying an outward oriented force perpendicular to the axis of variable spin (see Fig \ref{towercentri} right). The centrifuge consists of two $3$ $m$ long arms with an articulated platform at the end (Fig. \ref{towercentri} right), so that they swing outwards at increasing angular velocities of the centrifuge. The crate was mounted in one of the gondolas with a $37$ $kg$ counterweight on the other side. Power supply and data-connection of the crate was provided by means of a sliding contact at the axis of the centrifuge to the control room.

\begin{figure*}\label{g:centri}
\includegraphics[width=0.8\textwidth]{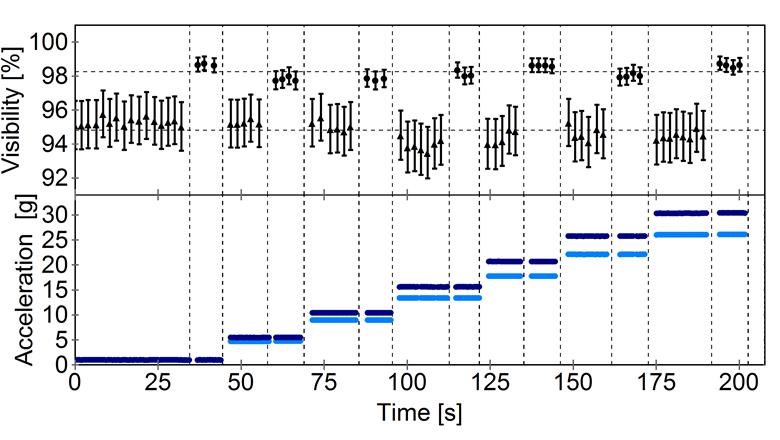} 
\caption{Centrifuge data: Graph displaying the time elapsed vs. the g-force and the visibility of the entangled photon state measured during the experiment in the DA ($\filledmedtriangleup$) and HV ($\bullet$) basis. The average of the visibilities is represented as horizontal dashed line ($94{.}8\%$ and $98{.}3\%$). Each of the points was calculated from $\approx 14{,}000$ coincident counts. The larger error bars in the DA measurements are the result of temperature fluctuations of the down-conversion crystal. The drop in visibility at $15$ $g$ is due to the effective cooling of the crystal during higher angular spinning speeds. The lower g-values (light blue) represent the measured data of the g-sensor at level 1 (detection) of the crate. Additionally the calculated acceleration at level 3 (source) is plotted (blue).}
\end{figure*}

Acceleration in the centrifuge experiment was varied by the angular velocity from $1$ $g$ to up to $30$ $g$ in approx. $5 g$ steps. The maximal tangential speed at the source layer is $174$ $km/h$. With respect to the stability of the optical setup, a movement of the UV-pump beam spot on the CCD camera of up to $2$ $pixel$ corresponding to $38$ $arcsec$ was monitored. This is well within the tolerances of the PMSM-fiber coupling lens system design. No drop in the count rates was observed. For each g value (set by the angular speed of the centrifuge) we evaluated the visibility for several minutes in two complementary basis states HV and DA (see Fig \ref{g:centri}). Note that a small reduction of the DA visibility stems from the fact that the temperature of the crystal was not stable at high accelerations. This was due to high wind speed of about $174$ $km/h$ which effectively cooled the system as indicated by the temperature sensor of the oven. 

\section{\label{sec:Discussion}Discussion}

We have conducted a micro-gravity (drop-tower) and a hyper-gravity (centrifuge) experiment, in which we used a genuine quantum mechanical system to evaluate whether the amount of entanglement varies at different accelerations. We implemented an entangled photon source together with a detection system and the required electronics in one single crate. We contiguously tested the quantumness of the system by measuring a witness which imposes a bound on the miminmum Bell-state fidelity, from the measured visibility of polarization correlations in two mutually unbiased measurement basis (HV and DA).

\begin{figure*}\label{g:total}
\includegraphics[width=0.8\textwidth, angle=0]{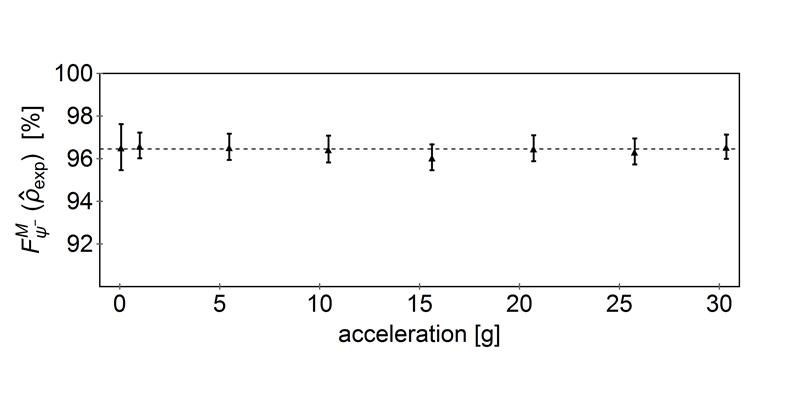} 
\caption{All data acquired during the experiments shown as the g-value vs. $F_{\Psi^-}^M (\hat{\rho}_{exp})= \frac{1}{2}(V_{HV}+V_{DA})$ ranging from $3$ $mg$ to up to $30$ $g$.
No deviation from the total average ($96{.}45\%$ represented as horizontal dashed line) for more than the estimated errors is visible.}
\end{figure*}

Fig. \ref{g:total} shows a summary graph displays the g-value vs. the entanglement fidelity measured for both conducted experiments in the falling tower and the centrifuge. A fidelity of above $96 \%$ at each acceleration level is achieved. Each data point was calculated using more than $10{,}000$ coincidence counts leading to an $3$ $\sigma$ error bar of $0{.}25 \% $. The measured fidelity is limited by numerous sources of  systematic errors, such as birefringence-induced transformations of the entangled photon state, temperature-dependent spectral characteristics of the SPDC source \cite{Steinlechner:2014}, as well as accidental coincidence counts \cite{Takesue:2010}.

Our experiment can rule out any hypothetical variation of entanglement to within the precision of the current measurement apparatus. The resolution in our experiment was limited by the short experimental time as well as residual drifts of the crystal-temperature and alignment issues, which are not correlated with the acceleration itself. The error bars shown in the graphs are calculated considering Poissonian statistics, as well as systematical errors due to temperature fluctuations. Due to insufficient temperature stabilisation of the crystal ($\pm 0{.}1^{\circ}C$) and gusts of cold air in the centrifuge, an additional error of $\pm 0{.}6\%$ was observed in the DA-Visibility. This corresponds to an uncertainty of $\pm  0{.}3\%$ in the Fidelity, in addition to statistical errors. The total error is well within the average fidelity taking all acquired data into account. 

In conclusion, we have used the techniques originally developed for quantum optics high precision experiments, to search for first experimental indications of unexpected relativistic and gravitational effects in quantum systems. Using the entangled system in the micro- and hyper-gravity condition experiments described above, we found that the phase and the entanglement of the state were not correlated to the acceleration of the system. These experiments therefore rule out any influence on the polarization entangled two-photon quantum state that could hypothetically cause a reduction in fidelity of more than $1{.}08\%$.

Our study tested photonic quantum entanglement in the case of flat space-time where the system undergoes uniform acceleration. Within this experiment we have shown that quantum entanglement should persist in a variety of gravitational settings ranging across the solar system.  

Such conditions would also accompany a rocket launch into space and are also relevant to subsequent quantum optics experiments carried out in space, under low-g conditions. We believe it is the first time that such a technology involving an entangled photon source was functional under these conditions and shown to be capable of withstanding the stresses. 

Our experiment demonstrates the extent to which state-of-the-art quantum hardware can be exposed to such harsh operational conditions and was intended to stimulate research on theories beyond the current paradigm which can be tested with the kind of experiments presented here. Our experimental platform represents a first test-bed that can readily be upgraded for measurements with higher precision, and higher-dimensional degrees of freedom, such as energy-time entanglement. We also envisage bringing such a system very close to zero-g conditions for several tens of seconds and reaching hyper-gravity environments of up to $150$ $g$ for many hours of operation.

\bibliography{sample}

We are especially thankful to Prof. Martin Tajmar and the Technical University Dresden (Germany) for their support in the drop-tower experiment. The centrifuge experiment was supported by Jens Schiefer and Clemens Greiner from AMST in Ranshofen (Austria). The Authors thank FFG-ALR (contract Nr. 844360), ESA (contract Nr. 4000112591/14/NL/US), FWF (P24621-N27) as well as the Austrian Academy of Sciences for their financial support.

\end{document}